# Pay No Attention to the Model Behind the Curtain

Philip B. Stark
Department of Statistics
University of California
Berkeley, CA 94720-3860
stark@stat.berkeley.edu
ORCID: 0000-0002-3771-9604

25 August 2022

To appear in *Pure and Applied Geophysics*

**Acknowledgments.**

I am grateful to Gil Anidjar, Monica Di Fiori, Robert Geller, Christian Hennig, Vladimir Kossobokov, Aaditya Ramdas, Jim Rossi, Andrea Saltelli, and Jacob Spertus for helpful suggestions.

**Abstract.**

Many widely used models amount to an elaborate means of making up numbers—but once a number has been produced, it tends to be taken seriously and its source (the model) is rarely examined carefully. Many widely used models have little connection to the real-world phenomena they purport to explain. Common steps in modeling to support policy decisions, such as putting disparate things on the same scale, may conflict with reality. Not all costs and benefits can be put on the same scale, not all uncertainties can be expressed as probabilities, and not all model parameters measure what they purport to measure. These ideas are illustrated with examples from seismology, wind-turbine bird deaths, soccer penalty cards, gender bias in academia, and climate policy.

> *reality ... what a concept.*
> —*Robin Williams*

**1. Introduction.**

There are reliable, empirically tested models for some phenomena. They are not the subject of this paper. There are also many models with little or no scientific connection to what they purport to explain.[1] Such 'ungrounded' models may be used because they are convenient,[2] customary, or familiar—even in situations where the phenomena obviously violate the assumptions of the models.

George Box famously said, "all models are wrong, but some are useful." (Box, 1976). This raises a number of questions, for instance: "useful for what?" "how can we tell whether a particular model is useful?"

Virtually any model is useful for getting a paper published or for rhetorical purposes—to persuade those who aren't paying close attention to technical details. This paper concerns utility for inference and policy.

A far better quotation from the same paper by Box is, "Since all models are wrong the scientist must be alert to what is importantly wrong. It is inappropriate to be concerned about mice when there are tigers abroad." (Box, 1976). This paper is about general issues that make models *importantly wrong.*

For the purpose of prediction, the proof of the pudding is in the eating: whether a model resembles reality matters less than whether the model's predictions are sufficiently accurate for their intended use. For instance, if a model predicts the fraction of customers who will be induced to buy by a marketing change accurately enough to decide whether to adopt that change, it doesn't matter whether the model is realistic in any sense. But for the purposes of explanation and causal inference—including formulating public policy—it is generally important for the map (the model) to resemble the real-world territory, even if it does not match it perfectly.

---

[1] I am ignoring trivial similarities, such as involving data of the same 'type' (e.g., real numbers, nonnegative numbers, integers, or categories) and I am ignoring heuristics (e.g., cyclical behavior or clustering). For example, see the discussion of the ETAS model below.
[2] One reason a model may be convenient is that software to fit or run the model is available. See Stark and Saltelli (2018).

The title of this chapter alludes to the 1939 film *The Wizard of Oz*. When Dorothy and her entourage meet The Wizard, he manifests as a booming voice and a disembodied head amid towering flames, steam, smoke, and colored lights. During this audience, Dorothy's dog Toto tugs back a curtain, revealing The Wizard to be a small, ordinary man at the controls of a machine designed to impress, distract, intimidate, and command authority. The Wizard's artificially amplified, booming voice instructs Dorothy and her entourage, "pay no attention to the man behind the curtain!"

The man behind the curtain is an apt metaphor for ungrounded models. Impressive computer results and quantitative statements about probability, risk, health consequences, economic consequences, etc., are often driven by a model behind the curtain—a model that we are discouraged from paying attention to. What claims and appears to be 'science' may be a mechanical amplification of the opinions and *ad hoc* choices built into the model, which lacks any tested, empirical basis. The choices in building the model—parametrizations, transformations, assumptions about the generative mechanism for the data, nominal data uncertainties, estimation algorithms, statistical tests, prior probabilities, and other "researcher degrees of freedom"—are like the levers in the Wizard's machine. And like the Wizard's machine, the models may have the power to persuade, intimidate, and impress, but not the power to predict or control.[3]

This paper tries to pull back the curtain to show how little connection there is between many models and the phenomena they purport to represent. It examines some of the consequences, including the fact that uncertainty estimates based on ungrounded models are typically misleading. Four ideas recur:

- *Quantifauxcation*. Quantifauxcation is a neologism for the common practice of assigning a meaningless number, then concluding that because the result is quantitative, it must mean something (and if the number has six digits of precision, they all matter). Quantifauxcation usually involves some combination of data, models, inappropriate use of statistics, and logical lacunae.
- *Type III errors*. A Type I error is a false positive, i.e., to reject a true hypothesis. A Type II error is a false negative, i.e., to fail to reject a false hypothesis. A Type III error is to answer the wrong question,[4] e.g., to test a statistical hypothesis that has little or no connection to the scientific hypothesis (an example of the *package deal* fallacy, whereby things that are traditionally grouped together are presumed to have an essential connection) (Stark, 2022). Another example is to conflate endpoints in clinical trials (e.g., to conclude that a treatment decreases mortality when there is only evidence that it decreases blood pressure).

---

[3] For a manifesto regarding responsible modeling, see Saltelli et al. (2020).
[4] Another definition of Type III error is "to draw the right conclusion for the wrong reason."

- *Models and Conspiracy Theories*. Models and conspiracy theories are technologies for providing a hard-to-falsify "explanation" for just about anything. Both purport to reveal the underlying causes of complicated phenomena. Both are supported by similar fallacies, including equivocation, changing the subject (red-herring arguments, *ignoratio elenchi*), appeals to ignorance (*argumentum ad ignorantiam*), appeals to inappropriate authority (*argumentum ad verecundiam*), ignoring qualifications (*secundum quid*), hasty generalization, *package deal*, cherry-picking facts, fallacy of the single cause, faulty analogies, and confirmation bias. "Science is Real," used as an unreflective slogan that considers "scientific" results to be trustworthy and credible—regardless of the quality of the underlying data and the methodology, conflicts of interest, and agenda—is emblematic.
- *Freedman's Rabbit Theorem.* There are two rabbit axioms:
  - The number of rabbits in any system is nonnegative.
  - For the number of rabbits in a closed system to increase, the system must include at least one rabbit (e.g., a pregnant female).

  From these, we conclude:

  **Proposition**: *You cannot borrow a rabbit from an empty hat, even with a binding promise to return the rabbit later.*

  **Theorem** (Freedman[5]): *You cannot pull a rabbit from a hat unless at least one rabbit has previously been placed in the hat.*

  Keeping the rabbit theorem in mind can help us tell whether results must depend heavily on assumptions. For instance, if model inputs are rates but model outputs are probabilities (the rabbit in this example), the model must assume that something is random, thereby putting the rabbit in the hat.

For the last 70 years or so, it has been fashionable to assign numbers to things to make them "scientific." Qualitative arguments are seen as weak, and many humanities and "soft" sciences such as history, sociology, and economics have adopted and exalted computation as scientific and objective.[6]

A 1978 opinion piece in *Nature* (Nature, 1978) put the issue well:

> It is objective and rational to take account of imponderable factors. It is subjective, irrational, and dangerous not to take account of them. As that champion of rationality, the philosopher Bertrand Russell, would have argued, rationality involves the whole and balanced use of human faculty, not a rejection of that fraction of it that cannot be made numerical.

Quantitative arguments and quantitative models require quantitative inputs. In many disciplines, the inputs, procedures, and models are all problematic. We now discuss some specific examples.

**2. Procrustes' quantifauxcation: forcing incommensurable things to the same scale.**

---

[5] Freedman, personal communication, circa 2001. He did not derive the theorem from axioms, but on more than one occasion told me, "to pull a rabbit from a hat, a rabbit must first be placed in the hat."

[6] For a defense of qualitative reasoning in science, see Freedman (2010b).

Procrustes of Greek mythology forced travelers to fit his bed, stretching them if they were shorter and cutting their limbs if they were taller. Such a brute-force reduction of disparate things to make them comparable is a common ingredient in quantifauxcation.

One common example is to form an "index" that combines a variety of things into a single number, for example, by adding or averaging "points" on Likert scales (Stark and Freishtat, 2014), doing arithmetic with numbers that represent disparate things by pretending that they are the same. This happens frequently in making composite scores from multivariate measurements, for instance, in university rankings.[7] The resulting outputs are somewhere between "difficult to interpret," "arbitrary," and "meaningless." This section gives some general and some specific examples related to cost-benefit analysis and the quantification of uncertainty, including using "probability" as a catch-all, blurring important differences.

## 2.1 'Utility' and cost-benefit analyses

It is often claimed that the only rational basis for policy is a quantitative cost-benefit analysis,[8] an approach tied to reductionism as described by Scoones and Stirling (2020. But if there is no rational basis for its quantitative inputs, how can relying on cost-benefit analysis be rational?

Not only are costs and consequences hard to anticipate, enumerate, or estimate in real-world problems, but behind every cost-benefit analysis is the assumption that all costs and all benefits can be put on a common, one-dimensional scale, typically money or abstract 'utility.' As a matter of mathematics, multidimensional spaces are not in general totally ordered: a binary relation such as "has more utility than" does not necessarily hold for every pair of points in the space. The idea that you can rank all aspects of a set of outcomes on the same one-dimensional scale is an *assumption*.

Luce and Tukey (1964) give conditions under which an individual's preferences among items with multiple attributes can be ranked on the same scale. The conditions (axioms) are nontrivial, as we shall see. Consider a collection of items, each of which has two attributes (e.g., a cost and a benefit). Each attribute has two or more possible values.

Consider sandwiches, for example:

| Attribute | Possible attribute values | | |
|---|---|---|---|
| **Filling** | peanut butter | turkey | ham |
| **Condiment** | mustard | grape jelly | cranberry sauce |

One of Luce and Tukey's conditions, *the (double) cancellation axiom*, requires:

> If you prefer peanut butter and jelly to turkey and cranberry sauce, and you prefer turkey and mustard to ham and jelly, then you *must* prefer peanut butter and mustard to ham and cranberry sauce.

---

[7] See also Desrosières (1998), chapter 3, "Averages and the Realism of Aggregates."
[8] For arguments for and against cost-benefit analysis, see Frank (2000). For a general treatment in the context of policy, see Sassone and Schaffer (1978).

For sandwiches, the cancellation axiom does not hold for everyone. It fails for me: sandwich type and condiment cannot be put on a single scale of 'utility' for me. Similar problems arise in more realistic cost-benefit analyses, not merely sandwiches, as a result of the Procrustean insistence on putting disparate qualitative dimensions on the same scale of 'utility' or money.

Luce and Tukey point out that the cancellation axiom does not always hold, and that when it fails, multidimensional attributes cannot be collapsed to a univariate scale. For instance, they write "one could test the cancellation axiom by examining a sufficiently voluminous body of ordinal data … and, thereby, test the primary ingredient in additive independence." (Luce and Tukey, 1964, p. 492).

Failure of the double cancellation axiom is not the only issue with cost-benefit analysis. To quantify costs in a cost-benefit analysis, in effect you must assign a dollar value to human life, including future generations; to environmental degradation; to human culture; to endangered species; and so on. You must believe that scales like "quality adjusted life-years" or "utility" are meaningful, reasonable, and a sound basis for decisions. Some scientists and philosophers are reluctant to be so draconian (Funtowicz and Ravetz, 1994).

Similarly, there is a slogan that *risk equals probability times consequences*. But what if the concept of probability doesn't apply to the phenomenon in question? (Some reasons that uncertainty may not be described well by probability are discussed below.) What if the consequences resist enumeration and quantification, or are incommensurable?

There is evidence that human preference orderings are not based on probability multiplied by consequences, i.e., on expected returns, losses, or 'utility'. For instance, there is a preference for "sure things" over bets: many people would prefer to receive $1 million for sure than to have a 10% chance of receiving $20 million, even though the expected return is double (Desrosiéres, 1998, p.49). And many people are loss-averse: they would prefer a 10% chance of winning $1 million over a 50% chance of winning $2 million with a 50% chance of losing $100 thousand, even though the expected return is 9.5 times larger. In a repeated game, basing choices on expectations might make sense, but in a single play, other considerations may dominate. Desrosiéres also quotes Leibnitz on the topic:

> As the size of the consequence and that of the consequent are two heterogeneous considerations (or considerations that cannot be compared together), moralists who have tried to compare them have become rather confused, as is apparent in the case of those who have dealt with probability. The truth is that in this matter—as in other disparate and heterogeneous estimates involving, as it were, more than one dimension—the dimension in question is the composite product of one or the other estimate, as in a rectangle, in which there are two considerations, that of length and that of breadth; and as for the dimension of the consequence and the degrees of probability, we are still lacking this part of logic which must bring about their estimate.[9]

Insisting on quantifying risk and on quantitative cost-benefit analyses requires doing things that may not make sense technically or morally. Moreover, I have yet to see a compelling example of incorporating uncertainty in the estimate of the probability (when the notion of 'probability' applies to the problem at all) and uncertainty in the consequences—much less the 'value' of those consequences. Such problems arise in probabilistic seismic hazard assessment (PSHA), discussed below.

---

[9] Desrosiéres, 1998, pp. 50–51, citing Coumet, 1970, in turn citing Leibniz, *New Essays on Human Understanding.*

## 2.2 Uncertainty

Just as costs and benefits cannot necessarily be put on the same scale, uncertainties cannot necessarily be put on the same scale. Nonetheless, many practitioners insist on using 'probability' as a catch-all to quantify all uncertainties.

*What is probability?*

Probability has an axiomatic aspect and a philosophical aspect. Kolmogorov's axioms, the mathematical basis of modern probability, are just that: mathematics. *Theories of probability* provide the philosophical glue to connect the mathematics to the empirical world, allowing us to interpret probability statements.[10]

The oldest interpretation of probability, *equally likely outcomes*, arose in the 16$^{th}$ and 17$^{th}$ centuries in studying games of chance, in particular, dice games (Stigler, 1986). This theory says that if a system is symmetric, there is no reason Nature should prefer one outcome to another, so all outcomes are "equally likely." For instance, if a vigorously tossed die is symmetric and balanced, there is no reason for it to land with any particular side on top. Therefore, all six possible outcomes are deemed "equally likely," from which many consequences follow from Kolmogorov's axioms as a matter of mathematics.

This interpretation of probability has trouble with situations that do not have the intrinsic symmetries of coins, dice, and roulette wheels. For example, suppose that instead of rolling a die, you toss a thumbtack. What is the probability that it lands point up versus point down? There is no obvious symmetry to exploit. Should you simply declare those two outcomes to be equally likely? And how might you use symmetry to make sense of "the probability of an act of nuclear terrorism in the year 2099?" That is even weirder. There are many events for which defining 'probability' using equally likely outcomes is unpersuasive or perverse.

The second approach to defining probability is the *frequency theory*, which defines probability in terms of limiting relative frequencies in repeated trials. According to the frequency theory, what it means to say 'the chance that a coin lands heads' is that if one were to toss the coin again and again, the fraction of tosses that resulted in heads would converge (in a suitable sense) to a limit; that limit is *defined* to be the probability that the coin lands heads. This theory makes particular sense in the context of repeated games of chance, because the long-run frequency that a bet pays off is identified to be its probability, tying the definition of probability to a gambler's long-run fortune.

There are many phenomena for which the frequency theory makes sense (e.g., games of chance where the mechanism of randomization is known and understood) and many for which it does not. What is the probability that global average temperature will increase by three degrees in the next 50 years? What is the chance there will be an earthquake with magnitude 8 or greater in the San Francisco Bay area in the next 50 years?[11] Can we repeat the next 50 years over and over to see what fraction of the time that happens, even in principle?

The *subjective theory* or (*neo-*)*Bayesian theory* may help in such situations. This theory defines probability in terms of degree of belief. According to the subjective theory, what it means to say "the probability that a coin lands heads is 50%" is that the speaker believes with equal strength that it will land heads as he or she believes that it will land tails. Probability thus measures the state of mind of the person making the probability statement.

---

[10] For a more technical discussion, see Freedman (2010a) and Stark and Freedman (2010); LeCam (1977). For an elementary discussion, see Stark (1997). For historical discussions, see Desrosiéres (1998) and Diaconis and Skyrms (2018).

[11] See Stark and Freedman (2010) for a discussion of the difficulty of *defining* "the chance of an earthquake."

The theory of equally likely outcomes is about the symmetry of the *coin*. The frequency theory is about what the *coin* will do in repeated tosses. The subjective theory is about what *I believe.* It changes the subject from geometry or physics to psychology. The situation is complicated further by the fact that people are not very good judges of what is going to happen, as discussed below. For making personal decisions—for instance, deciding what to bet one's own money on—the subjective theory may be a workable choice.

That said, LeCam (1977, pp. 134–135) offers the following observations:

> (1) The neo-Bayesian theory makes no difference between 'experiences' and 'experiments'.
>
> (2) It confuses 'theories' about nature with 'facts', and makes no provision for the construction of models.
>
> (3) It applies brutally to propositions about theories or models of physical phenomena with the same simplified logic which every one of us ordinarily uses for 'events'.
>
> (4) It does not provide a mathematical formalism in which one person can communicate to another the reasons for his opinions or decisions. Neither does it provide an adequate vehicle for transmission of information. (This, of course, is irrelevant to a purely 'personalistic' approach.)
>
> (5) The theory blends in the same barrel all forms of uncertainty and treats them alike.
>
> The above shortcomings belong to the neo-Bayesian *theory*. The neo-Bayesian *movement* has additional unattractive facets, the most important of which is its normative interpretation of the role of Statistics. Presumably a statistician who does not abide by its regulations is either irrational, inconsistent, or incoherent. …
>
> In summary, the Bayesian theory is wonderfully attractive but grossly oversimplified. It should be used with the same respect and the same precautions as the kinetic theory of perfect monoatomic gases.

Another approach to making sense of probability models, different from these three classical interpretations, is to treat probability models as *empirical commitments* (Freedman and Berk, 2010). For instance, coin tosses are not truly random: if you knew exactly the mass distribution of the coin, its initial angular velocity, and its initial linear velocity, you could predict with certainty how a tossed coin will land. But you might prefer to model the toss as random, at least for some purposes. Modelling it as random entails predictions that can be checked against future data—empirical commitments. In the usual model of coin tosses as fair and independent, all $2^n$ possible sequences of heads and tails in $n$ tosses of a fair coin are equally likely, which induces probability distributions for the lengths of runs, the number of heads, etc. It has been found that when the number of tosses is sufficiently large, the model does not fit adequately: data give evidence that the model is incorrect. In particular, there tends to be serial correlation among the tosses. And the coin is more likely to land with the same side up that started up.[12]

---

[12] The chance the coin will land with the same side on top that started on top is estimated to be about 51% by Diaconis et al. (2007). See also https://www.stat.berkeley.edu/~aldous/Real-World/coin_tosses.html, which reports an experiment involving 40,000 tosses (last visited 16 June 2022).

A final interpretation of probability is probability as *metaphor*. This seems to be how probability enters most policy applications and many scientific applications. It does not assert that the world truly behaves in a random way. Rather, it says that a phenomenon occurs 'as if' it were a casino game. Taleb (2007, pp. 127–129) discusses *the ludic fallacy* of treating all uncertainty as if it arose from casino games:

> "The casino is the only human venture I know where the probabilities are known, Gaussian (i.e., bell-curve), and almost computable." … [W]e automatically, spontaneously associate chance with these Platonified games. … Those who spend too much time with their noses glued to maps will tend to mistake the map for the territory. … Probability is a liberal art; it is a child of skepticism, not a tool for people with calculators on their belts to satisfy their desire to produce fancy calculations and certainties. Before Western thinking drowned in its "scientific" mentality, … people prompted their brain to think—not compute.

Probability as metaphor is closely tied to probability as an empirical commitment, although the key step of checking whether the model agrees adequately with the data is often omitted. We will see more on this below.

*Uncertainty and Probability*

Many scientists use the word 'probability' to describe anything uncertain. A common taxonomy classifies uncertainties as *aleatory* or *epistemic*. Aleatory uncertainty results from the play of chance mechanisms—the luck of the draw. Epistemic uncertainty results from ignorance. Epistemic uncertainty is 'stuff we don't know' but in principle could learn.

Canonical examples of aleatory uncertainty include coin tosses, die rolls, lotteries, radioactive decay, some kinds of measurement error, and the like. Under some circumstances, such things do behave (approximately) as if random—but generally not perfectly so, as mentioned above. Canonical examples of epistemic uncertainty include ignorance of the physical laws that govern a system or ignorance of the values of parameters in a system.

Imagine a biased coin that has an unknown chance $p$ of landing heads. Ignorance of the chance of heads is epistemic uncertainty. But even if we knew the chance of heads, we would not know the outcome of the next toss: it would still have aleatory uncertainty.

A standard way to combine aleatory and epistemic uncertainties involves using subjective (neo-Bayesian) prior probability to represent epistemic uncertainty. In effect, this puts individual beliefs on a par with unbiased physical measurements that have known uncertainties.

Calling two things by the same name does not make them the same. Combining aleatory and epistemic uncertainties by calling both 'probability' amounts to claiming that there are two equivalent ways to tell how much something weighs: I could weigh it on an actual physical scale or I could think hard about how much it weighs. The two are on a par. It claims that thinking hard about the question produces an unbiased measurement. As LeCam wrote, it does not distinguish between "experiments" and "experiences." Moreover, it implies that I know the accuracy of my internal 'measurement' from careful introspection. Hence, I can combine the two sources of uncertainty as if they are independent measurements of the same thing, both made by unbiased instruments.[13]

---

[13] When practitioners analyse complex systems such as climate, the economy, earthquakes, and such, the same observations they use as data in the problem are also the basis of their beliefs as reflected in the prior. But the analysis generally treats the data and prior as if they provided "independent" measurements—another fishy aspect of this approach.

This doesn't work. Combining uncertainties of entirely different sources and types by simply giving them the same name (probability) is another form of Procrustean quantifauxcation.

Sir Francis Bacon's triumph over Aristotle should have put to rest the idea that it is generally possible to make sound inferences about the physical world by pure ratiocination (some *Gedankenexperiments* notwithstanding[14]). Psychology, psychophysics, and psychometrics have shown empirically that people are bad at making even rough qualitative estimates, and that quantitative estimates are usually biased.

Moreover, the bias can be manipulated through processes such as *anchoring* and *priming*, as described in the seminal work of Tversky and Kahneman (1975). Anchoring, the tendency to stick close to an initial estimate, no matter how that estimate was derived, doesn't just affect individuals—it affects entire disciplines. The Millikan oil drop experiment to measure the charge of an electron (Millikan, 1913) is an example: Millikan's value was too low, supposedly because he used an incorrect value for the viscosity of air. It took about 60 years for new estimates to climb to the currently accepted value, which is about 0.8% higher (a small difference, but considerably larger than the error bars). Other examples include measurements of the speed of light and the amount of iron in spinach.[15] In these examples and others, somebody erred and it took a very long time for a discipline to correct the error because subsequent work did not stray too far from the previous estimate—perhaps because the first estimate made them doubt the accuracy of results that were far from it.

Tversky and Kahneman also showed that we are poor judges of probability, subject to strong biases from *representativeness* and *availability*, which in turn depends on the retrievability of instances, that is, on the vagaries of human memory. Their work also shows that probability judgments are insensitive to prior probabilities and to predictability, and that people ignore the regression to the mean effect—even people who have had formal training in probability. (Regression to the mean is the mathematical phenomenon that in a sequence of independent realizations of a random variable, particularly extreme values are likely to be followed by values that are closer to the mean.)

People cannot even accurately judge how much an object weighs with the object in their hands. The direct physical tactile measurement is biased by the density and shape of the object—and even its color.[16] The notion that one could just think hard about how global temperature will change in the next 50 years and thereby come up with a meaningful estimate and uncertainty for that estimate is preposterous. Wrapping the estimate in computer simulations of stylized, approximate physics distracts, rather than illuminates (Saltelli et al., 2015).

---

[14] A favorite example is Galileo's 'demonstration' that Aristotle was wrong that bodies of different masses fall at different rates: evidently, Galileo refuted Aristotle using a thought experiment. https://en.wikipedia.org/wiki/Galileo%27s_Leaning_Tower_of_Pisa_experiment (last accessed 1 June 2022)

[15] It is widely believed that spinach has substantially more iron than other green vegetables. This is evidently the result of a transcription error in the 1870s that shifted the decimal, multiplying the measured value by 10 (see, e.g., http://www.dailymail.co.uk/sciencetech/article-2354580/Popeyes-legendary-love-spinach-actually-misplaced-decimal-point.html0). That the original value was far too high was well known before the Popeye character became popular in the 1930s.

[16] E.g., Bicchi et al. (2008, section 4.4.3).

Humans are also bad at judging and creating randomness: we have *apophenia* and *pareidolia*, a tendency to see patterns in randomness.[17] And when we deliberately try to create randomness, what we make has fewer patterns than genuinely random processes would generate. For instance, we produce too few runs and repeats (e.g., Schulz et al., 2012; Shermer, 2008). We are over-confident about our estimates and predictions (e.g., Kahnemann, 2011; Taleb, 2007). And our confidence is unrelated to our actual accuracy (e.g., Krug, 2007; Chua et al., 2004).[18]

If I don't trust your internal scale or your assessment of its accuracy, why should your subjective (Bayesian) analysis carry any weight for me?[19]

In discussing the "neo-Bayesian" theory, LeCam (1977, pp. 155–156) gives these examples of uncertainty:

> It is clear that we can be uncertain for many reasons. For instance, we may be uncertain because (1) we lack definite information, (2) the events involved will occur according to the results of the spin of a roulette wheel, (3) we could find out by pure logic but it is too hard. The first type of uncertainty occurs in practically every question. The second assumes a well-defined mechanism. However, the neo-Bayesian theory seems to make no real distinction between probabilities attached to the three types. It answers in the same manner the following questions.
>
> (1) What is the probability that Eudoxus had bigger feet than Euclid?
> (2) What is the probability that a toss of a 'fair' coin will result in tails?
> (3) What is the probability that the $10^{137}+1$ digit of $\pi$ is a 7?
>
> Even Savage and de Finetti admit that, especially in cases involving the third kind of uncertainty, our personal probabilities are fleeting, more or less rapidly in that the very act of cogitating to evaluate precisely the probabilities is enough or can be enough to modify or totally overcome the uncertainty situation which one wanted to express.
>
> Thus, presumably, when neo-Bayesians state that a certain event *A* has probability one-half, this may mean either that he did not bother to think about it, or that he has no information on the subject, or that whether *A* occurs or not will be decided by the toss of a fair coin. The number ½ itself does not contain any information about the process by which it was obtained, fleetingly or not.
>
> As a final comment, it seem[]s necessary to mention that in certain respects the theory of personal probability is very similar to a theory of personal mass, which exhibits the same shortcomings.
>
> Suppose that a store owner is asked to assign weights to the items in his store. For this purpose he can group items in sets and compare them by hand. If a set *A* appears to him lighter than a set B we shall say that (A, B) ∈ R.[20] It is fairly easy to see, in the spirit of Theorem 3, that if the relation R is not compatible with an assignment of individual masses to the items and with the additivity of masses, the system is not very coherent. It is also possible to show that if there are enough items which could be indefinitely divided into 'equally weighty parts' the assignment of masses will be unique up to a multiplicative constant.

---

[17] https://en.wikipedia.org/wiki/Apophenia, https://en.wikipedia.org/wiki/Pareidolia (last accessed 16 November 2016)
[18] E.g., https://en.wikipedia.org/wiki/Overconfidence_effect (last accessed 16 November 2016)
[19] See Stark and Tenorio (2010).

Nobody would be particularly surprised however if it turned out that ten thousand peas which were judged all alike when compared pairwise turn out to be quite different when parted into two sets of 5000.

…

In spite of the theoretical possibility of assigning masses by hand comparison in this manner, nobody seems to claim that this is just what should be done in stores. Nobody even claims that since masses are masses there is no point in specifying whether they were obtained by hand comparison, or by using a spring scale or by using a balance.

…

If the process of measuring something as definite as masses by hand comparison seems rather unreliable, can one really expect a similar theory of measurement of ethereal opinions to inspire much confidence? If an indication of the process of measurement is helpful in the masses problem, it also appears necessary in the opinion problem.

Finally, an assignment of masses may conceivably be checked by experimenting with a scale, but the neo-Bayesian theory does not even pretend to make statements which could be checked by an impartial observer.

An editorial in *Nature* (1978) also pushes back on the idea that all risks can be quantified, much less quantified on the same scale:

> LORD ROTHSCHILD, speaking on British television last week, argued that we should develop a table of risks so we could compare, say, the risk of our dying in an automobile accident with the risk of Baader-Meinhoff guerillas taking over the nuclear reactor next door. Then we would know how seriously to take our risks, be they nuclear power, damage to the environment or whatever.
>
> …
>
> It is fine for Rothschild to demonstrate his agility with arithmetic, converting probabilities from one form to another (and implying that the viewers could not do it) but this is only the kindergarten of risk.
>
> …
>
> More than this, Rothschild confused two fundamental distinct kinds of risk in his table: known risks-such as car accidents-where the risk is simply calculated from past events; and unknown risks—such as the terrorists taking over a fast breeder—which are matters of estimating the future. The latter risks inevitably depend on theory. Whether the theory is a social theory of terrorism or a risk-tree analysis of fast breeder failure, it will be open to conjecture. And it ought to be remembered that the history of engineering is largely a history of unforeseen accidents. Risk estimates can be proved only by events. Thus it is easy for groups, consciously or unconsciously, to bend their calculations to suit their own objectives or prejudices. With unknown risks it is as important to take these into account as to come up with a number.

---

[20] This is rigorous way of writing a binary relation R (lighter than): the relation is a set of ordered pairs, the pairs of items that satisfy the binary relation.

In short, insisting on quantifying all kinds of uncertainty on the same scale—probability—is neither helpful nor sensible.

*Rates versus probabilities*

It is common to conflate empirical rates with probabilities. I have seen many examples in the literature of physics, geophysics (see below), medicine, and other fields. I am not saying that historical rates have no information about the process in question, but that historical rates are just that: historical rates. They are not probabilities, nor are they in general estimates of probabilities. Turning a rate into (an estimate of) a probability is pulling a rabbit from a hat. Unless the "physics" of the problem put probability into the hat, this is an illustration of Freedman's Rabbit Theorem.

Klemeš (1989) wrote eloquently about this false equivalence in hydrology:

> The automatic identification of past frequencies with present probabilities is the greatest plague of contemporary statistical and stochastic hydrology. It has become so deeply engrained that it prevents hydrologists from seeing the fundamental difference between the two concepts. It is often difficult to put across the fact that whereas a histogram of frequencies for given quantities … can be constructed for any function whether it has been generated by deterministic or random mechanism, it can be interpreted as a probability distribution only in the latter case. … Ergo, automatically to interpret past frequencies as present probabilities means *a priori* to deny the possibility of any signal in the geophysical history; this certainly is not science but sterile scholasticism.
>
> The point then arises, why are these unreasonable assumptions made if it is obvious that probabilistic statements based on them may be grossly misleading, especially when they relate to physically extreme conditions where errors can have catastrophic consequences? The answer seems to be that they provide the only conceptual framework that makes it possible to make probabilistic statements, i.e. they must be used if the objective is to make such probabilistic statements.

My experience in other branches of physical science and engineering is the same: equating historical rates with probabilities is so deeply ingrained that it can be impossible to get some people to see that there is any difference between the two.

Any finite series of dichotomous trials has an empirical rate of success. But the outcomes of a series of trials cannot tell you whether the trials were random in the first place. Suppose there is a series of Bernoulli trials,[21] that each trial has the same probability *p* of success, and that the trials are independent—like the standard model of coin tossing, treating 'heads' as 'success.' Then the Law of Large Numbers guarantees that the rate of successes converges (in probability) to the probability of success.

If a sequence of trials *is* random and the chance of success is the same in each trial, then the empirical rate of success is an unbiased estimate of the underlying chance of success. If the trials are random *and* they have the same chance of success *and* you know the dependence structure of the trials (for example, if the trials are independent), then you can quantify the uncertainty of that estimate of the underlying chance of success. *But the mere fact that something has a rate does not mean that it is the result of a random process.*

---

[21] A Bernoulli trial is a random dichotomous trial that can result in *failure* or *success,* generally represented numerically as 0 or 1, respectively.

For example, suppose a sequence of heads and tails results from a series of random, independent tosses of an ideal fair coin. Then the rate of heads will converge (in probability) to one half. But suppose I give you the sequence 'heads, tails, heads, tails, heads, tails, heads, tails, heads, tails, …' *ad infinitum*. The limiting rate of heads is ½. While that sequence *could* be the result of a sequence of fair random tosses, it is implausible, and it certainly need not be. Sequences of outcomes are not necessarily the result of anything random, and rates are not necessarily (estimates of) probabilities.

Here are two thought experiments:

1. You are in a group of 100 people. You are told that one person in the group will die next year. What is the chance it is you?
2. You are in a group of 100 people. You are told that one of them is named Philip. What is the chance it is you?

There is not much difference between these scenarios: both involve a rate of 1% in a group. But in the first one you are invited to say, 'the chance is 1%,' while in the second you are invited to say, 'that's a silly question.' The point is that a rate is not necessarily a probability, and that probability does not capture every kind of uncertainty.

In question 1, if the mechanism for deciding who will die in the next year is to shoot the tallest person, there is nothing random. There is no *probability* that you will be the person who dies—you either are or are not the tallest person, just as you either are or are not named 'Philip.' If the mechanism for deciding who will die is to draw lots and shoot whoever gets the short straw, that might be reasonably modeled as random, in which case the probability that you are the person who dies is indeed 1%. *The existence of a probability is in the method of selection, not in the existence of a rate.*

Everyday language does not distinguish between 'random,' 'haphazard,' and 'unpredictable,' but the distinction is crucial for scientific inference. 'Random' is a very precise statistical term of art. (The next section discusses how randomness might enter a scientific problem.)

Here is an analogy: to know whether a soup is too salty, a very good approach is to stir the soup thoroughly, dip in a tablespoon, and taste the contents of the tablespoon. That amounts to tasting a random sample of soup. If instead of stirring the soup I just dip the spoon in without looking, that would be a *haphazard* sample, a very different process. The second is *unpredictable* or *haphazard*, but it is not a random sample of soup, and it is not possible to quantify usefully the uncertainty in estimating the saltiness of the soup from a sample like that.

Notions such as probability, *P*-values, confidence intervals, etc., apply only if the data have a random component, for instance, if they are a random sample, if they result from random assignment of subjects to different treatment conditions, or if they have random measurement error. They do not apply to samples of convenience; they do not apply to haphazard samples; and they do not apply to populations. The mean and standard deviation of the results of a group of studies or models that is not a sample from anything does not yield actual *P*-values, confidence intervals, standard errors, etc. They are just numbers.

Rates and probabilities are not the same, and ignorance and randomness are not the same. Not all uncertainties can be put on the same scale.

*Probability Models in Science*

How does probability enter a scientific problem? It could be that the underlying physical phenomenon is random, as radioactive decay and other quantum processes are, according to quantum mechanics. Or it could be that the scientist deliberately introduces randomness, e.g., by conducting a randomized experiment or by drawing a random sample.

Probability can enter as a *subjective prior probability*. Suppose we want to estimate the probability $p$ that a particular coin lands heads (on the assumption that coin tosses are random). Surely $p$ is between zero and one. A common way to capture that constraint involves positing a *prior probability distribution* for $p$, for instance, assuming that $p$ was selected at random from the interval [0, 1], according to a uniform probability distribution. Unfortunately, positing any particular probability distribution for $p$ adds an infinite amount of information about $p$, information not contained in the constraint that $p$ is in the interval [0,1]. There are infinitely many probability distributions on the interval [0, 1], all of which satisfy the constraint: the constraint is not equivalent to a prior probability distribution, a function selected from an infinite-dimensional set of possibilities (see, e.g., Stark, 2015; Stark and Tenorio, 2010).

Theoretical results say that under some conditions, the prior does not matter asymptotically: the data eventually 'swamp' the prior as more and more observations are made. Those results involve conditions that might not hold in practice. In particular, it is not true in general for infinite-dimensional unknowns or for improper priors (see, e.g., Freedman, 1999). Nor is it necessarily true if the dimensionality of the problem grows as the number of data grows (Diaconis and Freedman, 1986). Nor is it true that nominally 'uninformative' (i.e., uniform) priors are actually uninformative, especially in high-dimensional spaces (Backus, 1987; Stark, 2015). For instance, consider a probability distribution that is uniform on the $d$-dimensional unit ball in Euclidean space. As $d$ grows, more and more of the mass is in an infinitesimal shell between radius 1-ε and 1.

Beyond the technical difficulties, there are practical issues in eliciting prior distributions, even in one-dimensional problems.[22] In fact, priors are almost never elicited—instead, priors are chosen for mathematical convenience or from habit. There are arguments in favor of the Bayesian approach from *Dutch book*: if you are forced to cover all possible bets and you do not bet according to a Bayesian prior, there are collections of bets where you are guaranteed to lose money, no matter what happens. According to the argument, you are therefore not rational if you don't bet in a Bayesian way. But of course, one is not forced to place bets on all possible outcomes (Freedman, 2010a). See section 2.2, above.

A fourth way probability can enter a scientific problem is through the invention of a *probability model* that is supposed to describe a phenomenon, e.g., a regression model, a Gaussian process model, or a stochastic PDE. But in what sense, to what level of accuracy, and for what purpose?[23] Describing data tersely and approximately (in effect, fitting a model as a means of data compression), predicting what a system will do next, and predicting what a system will do in response to a particular intervention are very different goals. The last involves causal inference, which is far more difficult than the first two (Freedman, 2010c).

---

[22] See, e.g., O'Hagan (1998). Typically, some functional form is posited for the distribution, and only some parameters of that distribution, such as the mean and variance or a few percentiles, are elicited.

[23] These problems may be worse in numerical modeling than in statistical modeling, yet they have received less systematic attention (Saltelli, 2019).

Fitting a model that has a parameter called 'probability' to data does not mean that the estimated value of that parameter estimates the probability of anything in the real world. Just as the map is not the territory, the model is not the phenomenon, and calling something 'probability' does not make it so.

Finally, probability can enter a scientific problem as *metaphor*: a claim that the phenomenon in question behaves 'as if' it is random. What 'as if' means is rarely made precise, but this approach is common, for instance, in stochastic models for seismicity.[24]

*Creating* randomness by taking a random sample or assigning subjects at random to experimental conditions is quite different from *inventing* a probability model or proposing a metaphor. The first may allow inferences if the analysis properly accounts for the randomization, if there are adequate controls, and if the study population adequately matches the population for which inferences are sought. But when the probability exists only within an invented model or as a metaphor, the inferences have little foundation. The inferences are no better than the assumptions. The assumptions and the sensitivity of the conclusions to violations of the assumptions have to be checked in each application and for each set of data: one cannot "borrow strength" from the fact that a model worked in one context to conclude that it will work in another context.

In summary, the word 'probability' is often used with little thought about why, if at all, the term applies, and many common uses of the word are rather removed from anything in the real world that can be reasonably described or modeled as random.

*Simulation and probability*

In some fields—physics, geophysics, climate science, sensitivity analysis, and uncertainty quantification in particular—there is a popular impression that probabilities can be estimated in a 'neutral' or 'automatic' way by doing Monte Carlo simulations: just let the computer generate the distribution.

For instance, an investigator might posit a numerical model for some phenomenon. The values of some parameters in the model are unknown. In one approach to uncertainty quantification, values of those parameters are drawn pseudo-randomly from an assumed joint distribution (generally treating the parameters as independent). The distribution of outputs is interpreted as the probability of various outcomes in the real world.

Setting aside other issues in numerical modeling, Monte Carlo simulation is a way to substitute computing for hand calculation. It is not a way to *discover* the probability distribution of anything; it is a way to estimate the numerical values that result from an *assumed* distribution. It is a substitute for doing an integral, not a way to uncover laws of Nature.

Monte Carlo doesn't tell you anything that wasn't already baked into the simulation. The distribution of the output comes from assumptions in the input (modulo bugs): a probability model for the parameters that govern the simulation. It comes from what you program the computer to do. Monte Carlo reveals the consequences of your assumptions, not anything new. The randomness is an assumption. The rabbit goes into the hat when you build the probability model and write the software. The rabbit does not come out of the hat without having gone into the hat first.

---

[24] See, e.g., Stein and Stein (2013), who claim that the occurrence of earthquakes and other natural hazards is like drawing balls from an urn, and make distinctions according to whether and how the number of balls of each type changes between draws. But why is the occurrence of natural hazards like drawing balls from an urn? This has no basis in physics.

Similarly, Gaussian process (GP) models (Kennedy and O'Hagan, 2001) are common in uncertainty quantification. I have yet to see a situation outside physics where there was reason to think that the unknown is a realization of a Gaussian process. The computed uncertainties based on GP models do not mean much if the phenomenon is not a realization of a GP.[25]

Uncertainty quantification and sensitivity analysis should acknowledge that the map is not the territory, that a histogram of model output is not a probability distribution, and that the sensitivity of model output to a parameter in the model does not translate automatically into sensitivity of the modelled system to that parameter.

*Cargo-Cult Confidence*[26]

Suppose you have a collection of numbers, for example, a multi-model ensemble of climate predictions for global warming,[27] or a list of species extinction rates for some cluster of previous studies. Take the mean (m) and the standard deviation (sd) of this list of numbers. Report the mean as an estimate of something. Calculate the interval [m-1.96sd, m+1.96sd]. Claim that this is a 95% confidence interval or that there is a 95% chance that this interval contains 'the truth.'

That is not a confidence interval for anything—and the probability statement is a further mangling of the interpretation of a confidence interval. If the collection of numbers were a random sample from some population and if that population had a Gaussian distribution (or if the sample size were large enough that you could invoke the central limit theorem), then the interval would be an approximate confidence interval for *something*. (The probability statement would still be garbled.)

But if the list is not a random sample or a collection of measurements with random errors, there is nothing stochastic in the data, and hence there can be no confidence interval. Performing confidence interval calculations on data that are not random is quantifauxcation. (See, e.g., van der Sluijs, 2016.)

A 95% confidence interval for a parameter is an interval calculated from a random sample using a method that—before the sample has been drawn—has at least a 95% probability producing an interval that includes the true value of the parameter. Once the sample has been drawn, everything is determined: even if the sample *was* random, the computed interval either does or does not include the true value of the parameter.

---

[25] However, see Neiswanger and Ramdas (2021) for a method of obtaining frequentist confidence sets from GP models when the prior is incorrect but the data are genuinely random.

[26] See Stark and Saltelli (2018). The title of this section alludes to Richard Feynman's discussion of *cargo-cult science* (Feynman, 1974). Cargo-cult confidence intervals involve calculations that look like confidence interval calculations, but they are missing a crucial element: the data are not a random sample. They involve implausible distributional assumptions as well.

[27] This is essentially what IPCC does.

Often, including in calculations I have seen in IPCC reports, people treat as if the *true parameter value* were random with a probability distribution centered at the estimate. This is something like Fisher's fiducial inference (Seidenfeld, 1992), which was virtually abandoned by statisticians many years ago. The treatment is backwards: if the *estimator* is unbiased, it is random with a probability distribution centered at the true, fixed, parameter value, not *vice versa*. Once the estimate has been made, it is also a fixed quantity.

## 3. Example: Probabilistic Seismic Hazard Assessment (PSHA)

Probabilistic seismic hazard analysis (PSHA) is the basis of seismic building codes in many countries. It is also used to help decide where to build nuclear power plants and nuclear waste disposal sites. PSHA purports to estimate the probability of a given level of ground shaking (acceleration), for instance, a level that would damage the containment structure. It involves modelling earthquakes as occurring at random in space, time and with random magnitude. Then it models ground motion as being random, conditional on the occurrence of an earthquake of a given magnitude in a given place.

From this, PSHA claims to estimate 'exceedance probability,' the chance that the acceleration in some particular place exceeds some tolerable level in some number of years. In the U.S., building codes generally require structures to withstand accelerations that will occur with probability of 2% or greater in 50 years.

PSHA arose from probabilistic risk assessment, which originated in aerospace and nuclear power, primarily. A big difference between PSHA and these other applications is that a spacecraft is an engineered system. Its properties are relatively well known (or predictable) even before humans had launched a manned spaceflight, as are those of the environment it is operating in, and so on. Even before a nuclear reactor was built, people knew something about nuclear physics and thermodynamics. They knew something about the physical properties of concrete and steel. They knew something about pressure vessels.

We know very little about earthquakes, other than their phenomenology. We don't really understand the physical generating processes (Geller et al., 2015). We don't know in detail how they occur. There is a big difference between an engineered system whose components can be tested and a natural system that is inaccessible to experimentation.

PSHA models earthquakes as a marked stochastic process with known parameters. The fundamental relationship in PSHA is that the probability of a given level of ground movement in a given place is the integral over space and magnitude of the conditional probability of that level of movement given that there is an event of a particular magnitude in a particular place times the probability that there is an event of a particular magnitude.

This is just the law of total probability and the multiplication rule for conditional probabilities, but where is the probability coming from? That earthquakes occur at random is an *assumption*, not a matter of physics. Seismicity is complicated and unpredictable: *haphazard*, but not necessarily *random*. The standard argument to calibrate the PSHA fundamental relationship requires conflating rates with probabilities. For instance, suppose a magnitude eight event has been observed to occur about once a century in a given region. PSHA would assume that, therefore, the chance of a magnitude 8 event is 1% per year.

That is wrong, for a variety of reasons. First, there is an epistemic leap from a rate to the existence of an underlying, stationary random process that generated the rate, as discussed above (see the quotation from Klemeš in particular). Second, there is an assumption that seismicity is uniform, which contradicts the observed clustering of seismicity in space and time. Third, this ignores the fact that even if seismicity were random and stationary, the historical rate is at best give an estimate of a probability, not the exact value of that probability.

Among other infelicities, PSHA conflates frequencies with probabilities in treating relationships such as the Gutenberg-Richter (G-R) law, the historical spatial distribution of seismicity, and ground acceleration given the distance and magnitude of an earthquake as probability distributions. For instance, the G-R law says that the log of the number of earthquakes of a given magnitude in a given region is approximately proportional to the magnitude. Magnitude is a logarithmic scale, so the G-R law says that the relationship between "size" and frequency is approximately linear on a log-log plot (at least over some range of magnitude). While the G-R law is empirically approximately true, PSHA involves the additional assumptions that the magnitudes of future earthquakes are drawn *randomly* from the G-R law.

PSHA relies on the metaphor that earthquakes occur as if in a casino game. According to the metaphor, it's as if there is a special deck of cards, the earthquake deck. The game involves dealing one card per time period. If the card is blank, there is no earthquake. If the card is an eight, there is a magnitude eight earthquake. If the card is a six, there is a magnitude 6 earthquake, and so forth.

There are tens of thousands of journal pages that, in effect, argue about how many cards of each kind are in the deck, how well the deck is shuffled, whether after each draw you replace the card in the deck and shuffle again before dealing the next card, whether you add high-numbered cards to the deck if no high card has been drawn in a while, and so on. The literature, and the amount of money spent on this kind of work, are enormous—especially given that it has been unsuccessful scientifically. Three recent destructive earthquakes were in regions that seismic hazard maps said were relatively safe (Stein et al., 2012; see also Panza et al., 2014; Kossobokov et al., 2015). This should not be surprising, because PSHA is based on a metaphor, not on physics.

Here is a different metaphor: earthquakes occur like terrorist bombings. We don't know when or where they're going to happen. We know that they could be be large enough to hurt people when they do happen, but not how large. We know that some places are easier targets than others (e.g., places near active faults), and that some are more vulnerable than others (e.g., places subject to soil liquefaction and structures made of unreinforced masonry). But there is no probability *per se*. We might choose to *invent* a probability model to try to improve law enforcement or prevention, but that is different from a generative model according to which terrorists decide when and where to strike by rolling dice. In principle, the predictions of such a model could be tested—but fortunately, such events are sufficiently rare that no meaningful test is possible.

What would justify using the casino metaphor for earthquakes? It might be apt if the physics of earthquakes were stochastic, i.e., truly random—but it isn't. It might be apt if stochastic models provided a compact, accurate representation of earthquake phenomenology, but they don't: the data show that the models are no good (see, e.g., Luen and Stark, 2012; Luen, 2010). The metaphor might be apt if the models led to useful predictions of future seismicity, but they don't (Luen, 2010).

A rule of the form, 'if there is an earthquake of a magnitude X or greater, predict that there is going to be another one within Y kilometers within Z days' predicts earthquakes quite well, without relying on stochastic mumbo jumbo pulled from a hat (Luen and Stark, 2008; Luen, 2010).

PSHA suffers from two of the issues we have been discussing, *viz.*, forcing all uncertainties to be on the same scale and conflating rates with probabilities. Cornell (1968), the foundational PSHA paper, writes:

> In this paper a method is developed to produce [various characteristics of ground motion] and their average return period for his site. The minimum data needed are only the seismologist's best estimates of the average activity levels of the various potential sources of earthquakes … The technique to be developed provides the method for integrating the individual influences of potential earthquake sources, near and far, more active or less, into the probability distribution of maximum annual intensity (or peak-ground acceleration, etc.). The average return period follows directly.
>
> …
>
> In general the size and location of a future earthquake are uncertain. They shall be treated therefore as random variables.

Whatever intuitive appeal and formulaic simplicity PSHA might have, what justifies treating everything that is uncertain as if it were random, with distributions that are known but for the values of a few parameters? Nothing. Moreover, the method does not work in practice (Mulargia et al., 2017). Note how many knobs and levers Cornell points out. Their settings matter.

I think we are better at ranking risks than quantifying them, and better at engineering calculations on human-built structures than we are at tectonic and seismological calculations. Thus, we might be better off asking questions by starting with a financial budget rather than a risk budget. Instead of asking, "how can we make this structure have a 90 percent chance of lasting 100 years?," we might be better off asking, "if we were willing to spend $10 million to harden this structure, how should we spend it?"

In summary, assigning numbers to things cannot always be done in a coherent or grounded way. To simply assume that the numbers must be meaningful because they are numbers is to commit quantifauxcation. In some situations, resisting the urge to assign numbers might be the wiser, safer, and more honest course.

## 4. Example: Avian-Turbine Interactions.

Wind turbine generators occasionally kill birds, in particular, raptors (Watson et al., 2018). This leads to a variety of questions. How many birds, and of what species? What design and siting features of the wind turbines matter? Can you design turbines or wind farms in a way that reduces avian mortality? What design changes would help?

I was peripherally involved in this issue for the Altamont Pass wind farm in the San Francisco Bay Area. Raptors are rare; raptor collisions with wind turbines are rarer. To measure avian mortality from turbines, people look for pieces of birds underneath the turbines. The data aren't perfect. There is background mortality unrelated to wind turbines. Generally, you don't find whole birds, you find bird fragments. Is this two pieces of one bird or pieces of two birds? Carcasses decompose. Scavengers scavenge. There are problems of attribution: birds may land some distance from the turbine they hit. How do you figure out which turbine is the culprit? Is it possible to make an unbiased estimate of the number of raptors killed by the turbines? Is it possible to relate mortality reliably to turbine design and siting?

The management hired a consultant, who modelled bird collisions with turbines as random, using a zero-inflated Poisson distribution[28] with parameters that depend parametrically on selected properties of the turbines. According to the model, the collisions are random and independent with a probability distribution that is the same for all birds. The expected collision rate follows a hierarchical Bayesian model that relates that rate parametrically to properties of the location and design of the turbine. The consultant introduced additional smoothing to make the parameters identifiable. In the end, he estimated the coefficients of the variables that the Poisson rates depend on—according to the model.

What does the model say? According to the model, when a bird approaches a turbine, in effect it tosses a biased coin. If the coin lands heads, the bird throws itself on the blades of the turbine. If the coin lands tails, the bird avoids the turbine. The chance the coin lands heads depends on some aspects of the turbine location and some aspects of its design, following a pre-specified formula that involves some unknown parameters. For each turbine location and design, every bird uses a coin with the same chance of heads, and the birds all toss the coin independently.

Whoever chose this model ignored the fact that some birds—including raptors—fly in groups: collisions are dependent. Why are avian-turbine interactions random? Why should they follow a Poisson distribution? Why do all birds use the same coin for the same turbine, regardless of their species, the weather, windspeed, and other factors? Why doesn't the chance of detecting a bird on the ground depend on how big the bird is, how tall the grass is, or how long it's been since the last survey? I don't know.

Perhaps the most troubling thing is that the formulation changes the subject from "how many birds does this turbine kill?" to "what are the numerical values of some coefficients in this zero-inflated Poisson regression model?" The analysis is a red herring, changing the subject from bird deaths to coefficients in the cockamamie model. It is also an example of a Type III error: testing a statistical model with little connection to the scientific question. Rayner (2012, p. 120) refers to this as *displacement*:

> Displacement is the term that I use to describe the process by which an object or activity, such as a computer model, designed to inform management of a real-world phenomenon actually becomes the object of management. Displacement is more subtle than diversion in that it does not merely distract attention away from an area that might otherwise generate uncomfortable knowledge by pointing in another direction, which is the mechanism of distraction, but substitutes a more manageable surrogate. The inspiration for recognizing displacement can be traced to A. N. Whitehead's fallacy of misplaced concreteness, 'the accidental error of mistaking the abstract for the concrete.'

Here is another example of a Type III error. Consider the statistical analysis of clinical trials. In the simplest setting, $N$ enrolled subjects meeting the trial criteria are assigned at random to treatment or control, for instance by selecting $n$ at random to receive the active treatment, and the remaining $N-n$ to receive a placebo. Then outcomes are observed. The scientific null hypothesis is that treatment does not affect the outcome, either subject-by-subject (the *strong null*) or on average (the *weak null*). A common way to test whether the treatment has an effect is using the 2-sample Student T-test. The statistical null hypothesis for that test is that the responses of the control group and of the treatment group are realizations of independent, identically distributed Gaussian random variables; equivalently, the two groups of responses

---

[28] This is a mixture of a point mass at zero and a Poisson distribution on the nonnegative integers.

are independent random samples from the same "superpopulation" of normally distributed values.

In the scientific experiment, the treatment and control groups are dependent because they arise from randomly partitioning a single group, so if a subject is in the treatment group, it is not in the control group, and *vice versa*. In the statistical null, the groups are independent. In the experiment, the only source of randomness is the random allocation of the $N$ subjects to treatment or control. In the statistical model, subjects' responses are random, as if the subjects were a random sample from a superpopulation. In the scientific experiment, there is no assumption about the distribution of responses for treatment or control. In the statistical model, responses (in the superpopulation) are Gaussian. What does testing the statistical null hypothesis tell us about the scientific null hypothesis?

Fisher (1935, pp. 50ff) himself notes:
> "Student's" $t$ test, in conformity with the classical theory of errors, is appropriate to the null hypothesis that the two groups of measurements are samples drawn from the same normally distributed population. … [I]t seems to have escaped recognition that the physical act of randomization, which, as has been shown, is necessary for the validity of any test of significance, affords the means, in respect of any particular body of data, of examining the wider hypothesis in which no normality of distribution is implied.

He goes on to describe how randomization by itself justifies permutation tests, which make no distributional assumptions about the data values *per se*, instead relying solely on the design of the experiment. While $P$-values computed for Student's $t$ test in this situation may be close to the $P$-values for a permutation test (under mild conditions, the former converges asymptotically to the latter), the hypotheses are conceptually very different. One is about the actual experiment and the randomization that was performed. The other is about imaginary superpopulations, bell curves, and independence. Using Student's $t$ test in this situation is to commit a Type III error. It also is a red herring argument, changing the subject from whether treatment has an effect to whether the data are plausibly IID Gaussian.

## 5. Example: Interruptions of academic job talks.

Women are disadvantaged in many ways in academia, including grant applications (Kaatz et al., 2014; Witteman et al., 2018), letters of reference (Schmader et al., 2007; Madera et al., 2009), job applications (Moss-Racusin et al., 2012; Reuben et al., 2014), and credit for joint work (Sarsons, 2015). Do academic audiences interrupt female speakers more often than they interrupt male speakers?[29] Blair-Loy et al. (2017) addressed this question by annotating 119 job talks from two engineering schools to note the number, nature, and duration of questions of various types, then fitting a zero-inflated negative binomial regression model with coefficients for gender, speaker's years since PhD, the proportion of faculty in the department who are female, and a dummy variable for university, and a dummy variable for department (CS, EE, or ME). The statistical hypothesis might be that the coefficient of gender in the "positive" model is zero (the negative binomial portion).

Blair-Loy et al. (2017) write:

---

[29] A separate question is, "and if so, does that make them less likely to be offered a job?"

> The standard choices for modeling count data are a Poisson model, negative binomial model, or a zero-inflated version of either of these models [55]. We prefer a zero-inflated, negative binomial (ZINB) model for this analysis …
>
> We now estimate ZINB models to address our first research question: do women get more questions than men during the job talk?

According to the ZINB model, questions occur as follows: in each talk, a biased coin is tossed. If it lands heads, there are no questions. If it lands tails, a (possibly) different biased coin is tossed repeatedly, independently, until that coin lands heads for the $k$th time. The number of questions asked is the number of tosses it takes to get the $k$th head. The probabilities that each coin lands heads and the value of $k$ are related parametrically to the covariates listed above.

Perhaps I lack imagination, but I can't see how that is connected to the number of questions in an academic job talk in the real world. It does produce a nonnegative integer, and the number of questions in a talk is a nonnegative integer. And it allows an arbitrarily large chance that the number of questions is zero, so it is more flexible than negative binomial regression without the extra mass at zero.

Blair-Loy et al. then test the hypothesis that the coefficient of gender in the "positive" part of the ZINB model is zero. That is, they ask whether it would be surprising for the estimated coefficient of gender to be as large as it was if interruptions of job talks followed the ZINB model but the true coefficient of gender in that model were zero. Setting aside the fact that the positive part alone does not reveal whether women or men get more questions on average,[30] what does that statistical hypothesis have to do with the original scientific question? The analysis changes the question from "are women interrupted more than men?" to "is the coefficient in a statistical model with no discernable scientific connection to the world zero?" This is an example of a Type III error, a red-herring statistical argument.

## 6. Example: Many analysts, one data set.

Silberzahn et al. (2018) involved 29 teams comprising 61 "analysts" attempting to answer the same question from the same data: are soccer referees more likely to give penalties ("red cards") to dark-skin-toned players than to light-skin-toned players. The 29 teams used a wide variety of models and came to different conclusions: 20 found a "statistically significant positive effect" and the other 9 did not.

This study was a great example of reproducible research, in that the data, models, and algorithms were made available to the public. Unfortunately, it was also a great example (29 great examples) of how *not* to model data and how to misinterpret statistical tests.

The teams used models and tests including:

- Least-squares regression, with or without robust standard errors or clustered standard errors, with or without weights
- Multiple linear regression
- Generalized linear models
- General linear mixed-effects models, with or without a logit link

---

[30] For instance, on average women might get more questions than men when they get questions at all (the "positive part" of the model), but they might not get any questions at all more often than men (the "zero model").

- Negative binomial regression, with or without a logit link
- Multilevel regression
- Hierarchical log-linear models
- Linear probability models
- Logistic regression
- Bayesian logistic regression
- Mixed-model logistic regression
- Multilevel logistic regression
- Multilevel Bayesian logistic regression
- Multilevel logistic binomial regression
- Clustered robust binomial logistic regression
- Dirichlet-process Bayesian clustering
- Poisson regression
- Hierarchical Poisson regression
- Zero-inflated Poisson regression
- Poisson multilevel modelling
- Cross-classified multilevel negative binomial regression
- Hierarchical generalized linear modeling with Poisson sampling
- Tobit regression
- Spearman correlation

The teams chose 21 distinct subsets of the 14 available covariates. The project included a phase of "round-robin" peer feedback on each team's model, after which the analytic approaches and models were generally revised. Then there was a second period of discussion and revision of the analyses. These "model-selection" steps alone would make the nominal significance levels of the tests wrong, even if all other aspects of the analyses were sound.[31]

With perhaps one exception, all the approaches involve positing parametric probability models for the issuance of red flags. Why is giving a red card random? Why does the

---

[31] To illustrate how tortuous the relationship between these models and the data is, consider one of the simpler models, logistic regression, which was used by two of the teams. Both teams used 6 covariates in their model, 4 of which were in both models. One of those teams used player's position (13 possible values, presumably represented as 12 dummy variables), height, weight, goals scored, victories, and the referee's country as the covariates, in addition to skin tone (which had 5 levels: 0, .25, .5, .75, and 1): 18 predictors of red cards. According to the logistic regression model, there is a single vector of coefficients ß common to all (player, referee) pairs. Each (player, referee) pair $i$ has its own logit-distributed random variable $U_i$; those logit variables are independent across pairs. Let $X_i$ denote the 18-vector of predictors for pair $i$. The referee gives the player a red card iff $X_i ß + U_i > 0$. See Freedman (2009, ch. 7).

Why is the issuance of a red cards independent across (player, referee) pairs, even though there are many players and typically several referees in each game? Why does the issuance of a red card depend on just those 18 predictors? Why does the threshold value of $U_i$ for issuing a red card depend linearly on those predictors, and with the same coefficients for all players?

issuance of a red card depend on just the predictors in the model? Why does it depend on those predictors in the way the model assumes?

The scientific hypothesis that referees give darker skin tone players more red cards has been turned into the statistical null hypothesis that red cards are issued according to a parametric probability model, and the coefficient of skin tone in that model is zero. The model changes the subject from what happens in the world to whether the value of a coefficient in an absurd cartoon model is zero. In particular, the *P*-value for the statistical hypothesis answers the question, *if red cards were issued according to the model with those covariates, and the "true" coefficient of skin tone is 0, what is the chance that the coefficient of skin tone would be estimated to be at least as large as it was estimated to be?*

If that is not how red cards are issued, the regression—and estimates of and hypothesis tests about coefficients in that regression—say little if anything about the world. This analysis is clearly a Type III error.

It is no wonder that the teams came to differing conclusions, since everything is simply invented from whole cloth, with no scientific basis for any of the models.

### 7. Example: Climate Models.

The IPCC treats all uncertainties as random:

> …quantified measures of uncertainty in a finding expressed probabilistically (based on statistical analysis of observations or model results or expert judgment).
>
> … Depending on the nature of the evidence evaluated, teams have the option to quantify the uncertainty in the finding probabilistically. In most cases, level of confidence…
>
> … Because risk is a function of probability and consequence, information on the tails of the distribution of outcomes can be especially important… Author teams are therefore encouraged to provide information on the tails of distributions of key variables…[32]

In these few sentences, the IPCC is asking or requiring people to do things that don't make sense and that don't work. As discussed above, subjective probability assessments (even by experts) are generally untethered to reality, and subjective confidence is unrelated to accuracy. Mixing measurement errors with subjective probabilities doesn't work. And climate variables have unknown values, not probability distributions.

Cargo-cult confidence confusion seems to be common in IPCC reports. For instance, the 'multi-model ensemble approach' involves taking a group of models, computing the mean and the standard deviation of their predictions, then treating the mean as if it were the expected value of the outcome (which it isn't) and the standard deviation as if it were the standard error of the natural process that is generating climate (which it isn't).[33] The resulting numbers say little if anything about climate.

---

[32] Mastrandrea et al. (2010) at p.2. The authors of the paper have expertise in biology, climatology, physics, economics, ecology, and epidemiology. To my surprise (given how garbled the statistics is), one author is a statistician.

[33] The IPCC also talks about simulation errors being 'independent,' which presupposes that such errors are random. https://www.ipcc.ch/publications_and_data/ar4/wg1/en/ch10s10-5-4-

## 8. Example: The Rhodium Group American Climate Prospectus.

The Bloomberg Philanthropies, the Office of Hank Paulson, the Rockefeller Family Fund, the Skoll Global Threats Fund, and the TomKat Charitable Trust funded a study[34] that purports to predict various impacts of climate change.

The report starts somewhat circumspect:

> While our understanding of climate change has improved dramatically in recent years, predicting the severity and timing of future impacts is a challenge. Uncertainty around the level of greenhouse gas emissions going forward and the sensitivity of the climate system to those emissions makes it difficult to know exactly how much warming will occur and when. Tipping points, beyond which abrupt and irreversible changes to the climate occur, could exist. Due to the complexity of the Earth's climate system, we don't know exactly how changes in global average temperatures will manifest at a regional level. There is considerable uncertainty…

But then the report makes rather bold claims:

> In this climate prospectus, we aim to provide decision-makers in business and government with the facts about the economic risks and opportunities climate change poses in the United States.

Yep, the 'facts.' They proceed to estimate the effect that climate change will have on mortality, crop yields, energy use, the labor force, and crime, *at the level of individual counties in the United States through the year 2099*. They claim to be using an 'evidence-based approach.'[35]

Among other things, the prospectus predicts that violent crime will increase just about everywhere, with different increases in different counties. How do they know? In some places, on hot days there is on average more crime than on cool days.[36] Fit a regression model to the increase.[37] Assume that the fitted regression model is a *response schedule*, i.e., it is

---

1.html (last accessed 16 November 2016) But modeling errors are *not* random—they are a textbook example of systematic error. And even if they were random and independent, averaging would tend to reduce the variance of the result, but not necessarily improve accuracy, since accuracy depends on bias as well.

[34] Houser et al. (2015). See also http://riskybusiness.org/site/assets/uploads/2015/09/RiskyBusiness_Report_WEB_09_08_14.pdf (last accessed 16 November 2016)

[35] To my eye, their approach is 'evidence-based' in the same sense that alien abduction movies are 'based on a true story.'

[36] Ranson (2014) claims, "[b]etween 2010 and 2099, climate change will cause an additional 22,000 murders, 180,000 cases of rape, 1.2 million aggravated assaults, 2.3 million simple assaults, 260,000 robberies, 1.3 million burglaries, 2.2 million cases of larceny, and 580,000 cases of vehicle theft in the United States."

[37] I used Ranson (2014) as the basis of a course project in 2018. The University of California, Berkeley, is in Alameda County, which has a population of about 1.68 million and an area of 739 square miles. The students split Alameda county into two pieces, east Alameda and west Alameda. They repeated Ranson's analysis in those two pieces and compared the results. The weather in east and west differed by an amount that is material and statistically significant.

how Nature generates crime rates from temperature. Input the average temperature change predicted by the climate model in each county; out comes the average increase in crime rate.

The hubris of these predictions is stunning. Think about the claims for a heartbeat. Even if you knew exactly what the temperature and humidity would be in every cubic centimeter of the atmosphere every millisecond of every day, you would have no idea how that would affect the crime rate in the U.S. next year, much less in 2099, much less at the level of individual counties. And that is before factoring in the uncertainty in climate models, which is enormous for globally averaged quantities (Regier and Stark, 2015), and even higher for individual counties. And it is also before considering that society is not a constant: changes in technology, wealth, and culture over the next 100 years surely matter, as evidenced by the recent rise in mass shootings (Ogasa, 2022).

Global circulation models (climate models) are theorists' tools, not policy tools: they might help us understand climate processes, but they are not a suitable basis for planning, economic decisions, and so on (see, e.g., Saltelli et al., 2015). The fact that intelligent, wealthy people spent a lot of money to conduct this study and that the study received high-profile coverage in *The New York Times* and other visible periodicals shows how effective quantifauxcation is rhetorically.

## 9. Discussion.

The reliance on models, especially for policy recommendations, has much in common with belief in conspiracy theories. Proponents of models and conspiracy theories generally treat agreement with a selected set of facts as affirmative evidence, even if there are countless competing explanations that fit the evidence just as well and evidence that calls their position into question. John von Neumann is reported (by Enrico Fermi) to have said, "With four parameters I can fit an elephant, and with five I can make him wiggle his trunk." Fit to past data says little about whether a model is 'correct' in any useful sense.

Proponents of models and conspiracy theories generally look for confirmation rather than for alternative explanations or critical experiments or observations that could falsify their position. Faced with inconvenient facts, modelers and conspiracy theorists tend to complicate their theories—e.g., by adding parameters to a model or agents to a conspiracy theory—rather than reconsider their approach. Proponents of models and conspiracy theories often borrow authority from other theories in a misuse of inductive reasoning.

For example, PredPol®, software for predictive policing, claims their model is reliable because it is used in seismology. It is the ETAS model (Ogata, 1988), which is rather problematic *within* seismology (e.g., Baranov et al., 2019; Grimm et al., 2022; Luen, 2010). Similarly, conspiracy theorists point to other conspiracy theories that have either been among the few that are true, or among the many that cannot be disproved, as support for their theory.

Models and conspiracy theories generally ride on unstated assumptions. For models, those assumptions often involve a version of the equivocation fallacy based on giving terms in the model the names of real-world phenomena. Labelling a term in the model "the coefficient of X" does not mean that its magnitude measures the effect of X in the real world.

And models often rely on authority or tribalism for support, rather than on empirical testing. For instance, there are entire disciplines built around a particular model, as econometrics is

---

The models for crime as a function of temperature differed by an amount that was statistically significant.

largely built on linear regression, seismic engineering is largely built on PSHA, and statistical earthquake prediction is largely built on ETAS.

Like conspiracy theories, models purport to reveal the underlying causes of complicated phenomena. Both offer an authoritative explanation of what we see, even when there are countless conflicting explanations that fit the data equally well. Both are well suited to reinforcing pre-existing beliefs through confirmation bias. Neither generally gets much "stress testing" (Mayo, 2018).

Conspiracy theories that contain a germ of truth are generally more successful: if some fact woven into the theory can be checked, that is seen as validating the theory, even if that fact is consistent with contradictory explanations. Similarly, models make some contact with data, lending the models credibility—even though it is typically possible to fit many contradictory models to the same data. There are strong social pressures to believe in models, despite gaps in logic, alternative explanations, and generally weak evidence, just like conspiracy theories. And like conspiracy theories, models are often created to support political positions or policies, and to question the models is discouraged by proponents of the policies, as is evident in the American Climate Prospectus (Rhodium Group, 2014).

van Prooijen and Douglas (2018) write:
> [C]onspiracy theories are *consequential* as they have a real impact on people's health, relationships, and safety; they are *universal* in that belief in them is widespread across times, cultures, and social settings; they are *emotional* given that [] emotions and not rational deliberations cause conspiracy beliefs; and they are *social* as conspiracy beliefs are closely associated with psychological motivations underlying intergroup conflict.

If the word "models" is substituted for "conspiracy beliefs" and "conspiracy theories," the paragraph remains true. Models are used to justify policy decisions with enormous societal impact; a broad spectrum of disciplines rely on them; they are used reflexively without adequate attention to whether they are appropriate or even tied to reality; they are used to justify political positions and there are strong social pressures—within science and more broadly in society—to believe model outputs. Models that purport to measure the cost or societal impact of climate change are a notable example (Saltelli et al., 2015).

Complexity is often a selling point of a model or conspiracy theory. For instance, if a computer model requires heroic supercomputing, that adds "weight" to the results. The American Climate Prospectus (Rhodium Group, 2014) proudly states:

> [This work] is only possible thanks to the advent of scalable cloud computing. All told, producing this report required over 200,000 CPU-hours processing over 20 terabytes of data, a task that would have taken months, or even years, to complete not long ago.

Similarly, if a conspiracy theory involves more colluders and evil actors, it may be more appealing.

Tribalism is often a factor. People connected by religious and political affiliations tend to believe the same conspiracy theories. The same is true for models—both within particular disciplines,[38] and under the general rubric of "science is real," suggesting that because models are "scientific," they are credible and trustworthy.

Like conspiracy theories, models are often promulgated by people who do not even understand the theory: the fact that it is "science" is enough.

Another similarity is that, like conspiracy theories, models can be nearly impossible to disprove, because so many models are consistent with the observations. By definition, underdetermined problems admit many solutions that can differ substantially from each other, even though all fit the data equally well.

Gorvett (2020) summarizes some characteristics of successful conspiracy theories: convincing culprits, collective anxieties, tribalism, providing certainty amidst uncertainty, and exploiting knowledge gaps. The parallels with successful—but scientifically bogus—models are striking.

In closing, I quote the five principles of Saltelli et al. (2020) to help ensure that models serve society:
- Mind the assumptions: assess uncertainty and sensitivity.
- Mind the hubris: complexity can be the enemy of relevance.
- Mind the framing: match purpose and context.
- Mind the consequences: quantification may backfire.
- Mind the unknowns: acknowledge ignorance.

**References**.

---

[38] Almost the entire field of econometrics is regression models, even though the assumptions of regression are rarely met in economic applications. See, e.g., Kennedy (2001).